\documentstyle[twocolumn,aps]{revtex}
\begin{document}
\draft

\twocolumn[\hsize\textwidth\columnwidth\hsize\csname @twocolumnfalse\endcsname
\title{
Phase separation of crystal surfaces: a lattice gas approach
}
\author{
Joel D. Shore and Dirk Jan Bukman
}
\address{
Department of Physics, Simon Fraser University,
        Burnaby, British Columbia, Canada V5A 1S6
}
\date{\today}
\maketitle
\begin{abstract}
We consider both equilibrium and kinetic aspects of the phase separation
(``thermal faceting") of thermodynamically unstable
crystal surfaces into a hill--valley structure.  The model we study is an
Ising lattice gas for a simple cubic crystal with nearest--neighbor attractive
interactions and weak next--nearest--neighbor repulsive interactions. It
is likely applicable to alkali halides with the sodium chloride structure.
Emphasis is placed on the fact that the equilibrium crystal shape can be
interpreted as a phase diagram and that the details of its structure tell us
into which surface orientations an unstable surface will decompose.
We find that, depending
on the temperature and growth conditions, a number of interesting behaviors
are expected.  For a crystal in equilibrium with its vapor,
these include a low temperature regime with logarithmically--slow separation
into three symmetrically--equivalent facets, and a higher temperature regime
where separation proceeds as a power law in time
into an entire one--parameter family of surface orientations.  For a crystal
slightly out of equilibrium with its vapor (slow crystal growth or etching),
power--law growth should be the rule at late enough times.  However, in the
low temperature regime, the rate of separation rapidly decreases as the
chemical potential difference between crystal and vapor phases goes to zero.
\end{abstract}

\vskip2pc]

\section{Introduction}\label{sec1}

The decomposition (or ``faceting") of a surface into pieces of surfaces of
other orientations is a problem of long--standing interest in the materials
science and physics communities.
It can occur in a variety of circumstances, including when a crystal is being
grown or etched \cite{ExperimentalLiterature},
when deposition is occurring through such processes as
molecular beam epitaxy \cite{MBE,SiegertPlischke}, when other materials are
adsorbed onto the surface\cite{ChemicalAbsorption},
or when an electric current is applied across the sample
\cite{Electromigration,KrugDobbs}.  However, such external perturbations are
not always necessary to cause this decomposition:
If the surface tension is sufficiently anisotropic, some surface
orientations will be thermodynamically unstable\cite{Herring}.
Then, a surface which is initially prepared in such an
orientation will spontaneously decompose into a faceted structure, evolving
towards a final equilibrium configuration consisting of large facets of
stable orientations.
This decomposition has been variously referred to as ``equilibrium faceting,"
``thermal faceting," ``thermal etching,'' ``Herring reconstruction,'' or
``hill-valley reconstruction''
\cite{ExperimentalLiterature,Herring,Knoppik,Mullins}.

Recently, there has been a resurgence of interest in this process in both the
experimental\cite{Phaneuf1,Phaneuf2,Yoon} and theoretical physics communities
\cite{OurLongPaper,Stewart,Fong,Vlachos}.  In the theoretical community, the
close analogy to phase separation of binary liquids or alloys has been
pursued.  Hence, the process has been dubbed ``phase separation" or ``spinodal
decomposition" of crystal surfaces \cite{Stewart,Fong}.  (The terms
``faceting,"
``phase separation," ``decomposition," and ``coarsening" each have their
drawbacks in describing this process in all its various manifestations but,
lacking better terminology, we will use all these terms and use them more or
less interchangeably.)

The process, which is illustrated in Fig.\ \ref{surface}, proceeds as follows:
The surface, initially prepared in an unstable orientation, first decomposes
into small
pieces (``facets") of stable surface orientations, in spite of the fact that
this increases the total surface area.  The faceted surface then coarsens
over time in order to minimize the energy associated with the edges between
the different surface orientations.
The dynamical process by which the size of the facets grows over time is indeed
closely analogous to the phase separation of a binary
liquid or alloy, and recent theoretical work has attempted to elucidate
the analogy and to investigate the dynamics associated with such
separation \cite{OurLongPaper,Stewart,Fong,Vlachos}.
Much earlier work along
these lines was performed by Mullins\cite{Mullins}, and he predicted power law
growth of the facet size $L$ with time $t$: $L(t) \sim t^n$ with
$n = 1/4$, $1/3$, or $1/2$ for the mass-transfer mechanisms of
surface--diffusion, volume--diffusion, or evaporation--condensation,
respectively.  However, the experimental situation has remained
murky.  Generally, it has been found that such faceting occurs only very
slowly ({\it e.g., } $n \approx 0.1$) if at all under near-equilibrium
conditions (see discussion in \cite{Vlachos}),
and indeed, some have argued that such faceting is not even a thermodynamic
phenomenon but occurs only under driven conditions of crystal growth or
etching\cite{ExperimentalLiterature}.  Nonetheless, it is known theoretically
that such faceting should in principle occur in equilibrium if the surface
tension is
sufficiently anisotropic and some of the experimental observations do
support this \cite{Knoppik,Heyraud}.  Thus we are led to conclude that
one should seek a dynamical explanation to understand why so little faceting
is seen\cite{Elastic}.

Two of the recent studies of the dynamics have used solid--on--solid
approximations of an Ising lattice gas on a cubic lattice with interactions
appropriate for alkali halide materials such as sodium chloride
(NaCl)\cite{RottmanWortis,ShiWortis}.
Shore, Holzer, and Sethna\cite{OurLongPaper} considered the special case of
a [111] surface and argued that the coarsening of the surface should be only
logarithmic with time (at long times) for quenches to all temperatures $T$ at
which the [111] surface is unstable.  This claim was supported by their
Monte Carlo simulations.  Vlachos, Schmidt, and Aris\cite{Vlachos}
performed Monte Carlo simulations for the coarsening of $[hk0]$
surfaces, concentrating in particular on a [210] surface.  Although they draw
no firm conclusions about the asymptotic growth laws,
we believe that the results presented there---of
anomalously low exponents which decrease with decreasing temperature---are
compatible with (although certainly not proof of) logarithmic growth at long
times \cite{OurLongPaper,GrestAndSrolovitz}.

The approach of Stewart and Goldenfeld\cite{Stewart}
and Liu and Metiu\cite{Fong} has been along
somewhat different lines, closer in spirit to the original work of Mullins.
They have looked at the problem in the continuum
limit, which is generally claimed to be valid when one is interested in
behavior at long times and on long length scales.  In this limit, the driving
force for the initial breakup of the surface into facets is provided by a
sufficiently anisotropic surface tension.
Also included in the free energy functional
is a term which suppresses rapid changes in the surface normal\cite{Fong}.
Such a term has two effects:  First it rounds off the edges and corners on
the surface, thus preventing the occurrence of singularities, which would lead
to divergences in the Langevin equations.  Second, it
reproduces, in a macroscopic sense, the energy costs associated with edges
and corners, thus providing the driving force for the coarsening of the surface
structure.  For this model, Liu and Metiu concluded that coarsening proceeds
logarithmically for quasi-one-dimensional ordering.  Simulations of the
resulting Langevin equation for a two-dimensional surface
gave $L(t) \sim t^n$ with $n \approx 0.13$ and $0.23$ for
surface diffusion and for evaporation--condensation mechanisms, respectively.
These results were noted to be in fairly good agreement with simple
power--counting arguments on the Langevin equation, which suggest
$n = 1/6$ and $1/4$ for the two different mechanisms.

Although continuum approaches are a powerful tool, and currently enjoy great
favor in the study of kinetics of growth, they suffer from at least two major
limitations that make it useful to also consider microscopic ({\it e.g., }
lattice gas) models.  The first is that, since they deal
with the system on a coarse--grained level, one generally needs to assume some
form for the input parameters, in this case the orientation--dependence of
the surface tension.  Thus, microscopic models are useful to give us
guidance on what form to choose for the surface tension.  Indeed, in Sec.\
\ref{sec3b} we will find
that the form of the surface tension that arises in one such microscopic model,
and may occur quite generally, is considerably different from that which
one might have expected from naive considerations.

The second limitation is that such models, at least as currently formulated,
appear to be unable to properly model aspects of the problem where the
atomic discreteness is fundamental and cannot be ignored.  This leads to the
conflicting predictions discussed above (logarithmic vs.\ power--law growth)
between the microscopic and continuum approaches for the coarsening of an
unstable $[111]$ surface of NaCl.

The lattice gas models, of course, have their own limitations
({\it e.g.,\ } the neglect of elastic effects \cite{Elastic}), but
at this point a fairly complete picture of the equilibrium and
kinetic aspects of the faceting process within a microscopic model would be
useful in advancing the understanding of the faceting process in the real
world.
In this paper, we will try to present such a unified picture of faceting
within the context of the Ising lattice gas model for NaCl.  The picture
combines and generalizes some of our earlier work on the faceting of a [111]
surface\cite{OurLongPaper}, and some more recent results concerning the phase
diagram of the six--vertex model\cite{SixVertex}.  We also include
a discussion of how the kinetics of the faceting process will be altered for
the case of a driven interface, that is, one in which the crystal is being
slowly grown or etched.

The outline of the paper is as follows:  In Section II, we introduce the
Ising lattice gas model which we will study, discuss the general
features of its equilibrium crystal shape (ECS), and explain how the ECS tells
us which surface orientations are unstable and into what surfaces such an
unstable orientation will decompose.  In Section III, we consider in detail
the equilibrium and kinetic aspects of the phase separation in two different
temperature regimes.
In Section IV, we discuss how the kinetics is altered in the case when the
crystal is no longer in equilibrium with its vapor (growth or etching).
Finally, in Section V, we present a summary of our conclusions.

\section{Equilibrium crystal shape}\label{sec2}

The problem of determining the equilibrium shape of a crystal by minimizing
its total surface free energy
subject to a constraint of fixed volume has been
considered for over a century \cite{EarlyWork}.  The major result is the
celebrated Wulff construction by which one determines the ECS given
the surface free energy (surface tension) as a function of the surface
orientation.

Only much more recently, however, has the ECS of some simple microscopic models
for crystals been considered
\cite{RottmanWortis,Wortis,VanBeijeren,JayaprakashSaam}.
Of particular interest to us is the Ising lattice gas on a simple cubic lattice
with nearest--neighbor attractive interactions and next--nearest neighbor
repulsive interactions, first studied by Rottman and Wortis
\cite{RottmanWortis}.  The Hamiltonian is
\begin{equation}
        {\cal H} = -J_1 {\sum_{\rm NN} s_i s_j} + J_2 {\sum_{\rm NNN} s_i s_j}
		+ {\Delta\mu\over{2}} {\sum_i s_i}
        \ ,
        \label{hamiltonian}
\end{equation}
where the $s_i$ take on the values $+1$ and $-1$.  In the lattice gas context,
$+1$ is taken to represent an atom occupying the site and $-1$ is taken to
represent an empty site.  The first sum is over
all pairs of nearest--neighbors (NN) while the second is over all pairs of
next--nearest--neighbors (NNN).  We have chosen our sign convention so that
both $J_1$ and $J_2$ are positive when the NN bonds are
attractive and the NNN bonds are repulsive.  We are interested in the case
where $J_1/J_2 > 4$, in which case the ground state is (using spin language)
``ferromagnetic" \cite{OurLongPaper}.

In many instances, it will be useful to consider an interface between occupied
and unoccupied sites ({\it i.e., } between solid and gas)
in the [111]--restricted--solid--on--solid (RSOS) approximation
\cite{ShiWortis,OurLongPaper}.
This approximation can be obtained by viewing the interface from the [111]
direction and requiring that none of the interface be hidden from view by other
parts of the interface (see Fig.\ \ref{surface}).  It is equivalent to taking
the limit $J_1/J_2 \to \infty$ with $T/J_2$ fixed.
Configurations in this model can also be viewed as tilings of the plane
by ${60^\circ}$ rhombi
of three different orientations.  In this representation, the energetics are
reproduced by assigning an energy of ${2 J_2}$ to each border between
unlike rhombi, and the coarsening process involves the phase separation of the
three types of rhombi.

Finally, $\Delta\mu \equiv \mu_{\rm s} - \mu_{\rm v}$ represents a difference
in chemical potential between solid and vapor.  (In spin language it would be
a magnetic field.)
For the remainder of this paper, with the exception of Section~\ref{sec4},
we will be considering the case in which the solid and vapor
are in equilibrium ($\Delta\mu = 0$).

The model of Eq.\ (\ref{hamiltonian}) and its RSOS approximation
were proposed\cite{ShiWortis,Wortis}
to represent materials such as sodium chloride (NaCl), where
the ions of different species (and opposite charge) would have an attractive
interaction while those of the same species would repel one
another\cite{RoughlySpeaking}.  The ECS for this model is
show in Fig.\ \ref{ECS}.  It was found \cite{RottmanWortis} that the ECS
remains
strictly cubical, with (macroscopically) sharp edges and corners up to the
corner--rounding transition at a temperature $T=T_{\rm CR}$, at which point the
crystal first rounds at the corners.  For the RSOS
model, $T_{\rm CR}$ can be calculated exactly \cite{ShiWortis,OurLongPaper} and
is given by \cite{TemperatureMeasurement}
\begin{equation}
        T_{\rm CR} = {-4 J_2\over{\ln [1/3 -
                        5/(9\alpha^{1/3})
                        + \alpha^{1/3}]}} \; \approx \; 7.1124\ldots J_2
        \ ,
        \label{TCR}
\end{equation}
where $\alpha \equiv {1\over{6}} ( {11\over{9}} + \sqrt{23/3})$.
As the temperature is further increased,
the rounded region spreads out along the edges until the edge--rounding
temperature $T_{\rm ER}$, at which point the entire edge becomes rounded.
The smooth $\{100\}$ facets still remain, however, up to the roughening
temperature $T_{\rm R}$ \cite{CrystalographicNotation}.
Finally, above $T_{\rm R}$, the entire crystal shape is rounded.

We make two further observations about these interfacial phase transitions.
The first is that when $J_2 = 0$ or is attractive, then
$T_{\rm CR} = T_{\rm ER} = 0$.
Therefore the corner and edge rounding transitions are the result of the
competing attractive and repulsive interactions.  The second is that within
the [111]--RSOS model, $T_{\rm ER}$ and $T_{\rm R}$ are infinite.
This is a result
of the fact that setting $J_1 \to \infty$ suppresses the fluctuations which
are responsible for edge rounding and the roughening of the facets.
Thus, the usefulness of the RSOS approximation is restricted to temperatures
$T < T_{\rm ER}$.  It is most reliable for surface orientations close to
$[111]$.

Experimentally, in NaCl, the corner--rounding transition is found to
occur at a temperature $T_{\rm CR}\approx 920$ K
\cite{Heyraud,ShiWortis,Wortis,RoughSense}.
The sharp cubical shape and the shape with facets but no
sharp corners or edges have both been observed
experimentally \cite{Heyraud}.  The intermediate shape with sharp edges
but rounding at the corners was also seen on some crystals.  However, it is
likely that these crystals were not sufficiently equilibrated, so one would
have to say that there is, as of yet, no experimental confirmation of this
intermediate shape, and thus, of the claim that the corner-- and edge--rounding
temperatures are distinct (See \cite{Heyraud} and the discussion in
\cite{Wortis}).

It is a theorem due to Herring \cite{Herring} that those surface
orientations which do not appear on the ECS are
thermodynamically unstable (and will thus break up into pieces of
surface of stable orientations).  If we represent each surface orientation
by a unit vector normal to that surface, then we can use a unit sphere to
represent the various surface orientations.  For the model specified by
Eq.\ (\ref{hamiltonian}), Fig.~\ref{AllowedOrientations}
shows an eighth of such a unit sphere, with shading on the sphere used to
indicate those surface orientations which are stable in various temperature
regimes.  Since only the $\{100\}$ facet orientations appear on the crystal
shape (Fig.\ \ref{ECS}) for $T < T_{\rm CR}$, they are the only stable surface
orientations in this temperature regime.
At $T_{\rm CR}$, the $[111]$ surface orientation (which is at
the center of our diagrams in Fig.\ \ref{AllowedOrientations}) and all those
orientations on the lines connecting it to the three nearest facet orientations
become (marginally) stable.  As the temperature is raised further,
more and more surface orientations become stable.
However, it is not until the temperature reaches $T_{\rm ER}$ that all
surface orientations appear on the ECS and are thus stable.

Herring's theorem is actually a specific consequence of a more general fact:
The ECS is essentially a phase diagram for the surface
orientations \cite{Andreev,Wortis,SixVertex}.  More precisely, the height of
the ECS above a reference plane, $F(h,v)$, as a function of the two in-plane
coordinates $(h,v)$, can also be considered as a free energy as a
function of ``fields" $(h,v)$ \cite{ConfusingNotation}.
These fields are conjugate to the two--component order parameter
$(x,y)$ specifying the surface orientation.
The projection of the surface tension onto the reference plane is then given
by the free energy as a function of surface orientation $F(x,y)$.
As is usually the case, the conjugacy
of the fields and order parameters means that these two free energies
are related by a (two--dimensional) Legendre transformation \cite{Andreev}:
\begin{equation}
F(h,v) = \min_{x,y} \, \lbrace F(x,y) - hx - vy \rbrace,
\label{Legendre}
\end{equation}
which implies $F(h,v) = F(x,y) - hx - vy$ with $x$ and $y$ chosen such that
${\partial F(x,y)/\partial x} = h$
and ${\partial F(x,y)/\partial y} = v$.  Equivalently, we can write
$x = -{\partial F(h,v)/\partial h}$ and $y = -{\partial F(h,v)/\partial v}$,
which simply states the fact that the surface orientation is specified by the
slope on the crystal shape $F(h,v)$.

Because of the mapping between the ECS and a free energy surface,
we can transfer all our terminology and knowledge about phase
diagrams over to the realm of equilibrium crystal shapes!  For example,
sharp edges in the ECS are places where $F(h,v)$ has a discontinuity in the
first derivatives,
{\it i.e.,} first--order phase transitions; whereas boundaries where a facet
smoothly joins the rough regions of the crystal are places where the
first derivatives of $F(h,v)$ are continuous but higher--order derivatives are
not, and they are thus
considered second--order transitions.  Even the scaling behavior at these
boundaries can be analyzed using the tools of critical phenomena
\cite{JayaprakashSaam,PT}.

At first--order lines (``edges"), one has coexistence between the two ``phases"
(surface orientations).  The corners of the ECS for $T < T_{\rm CR}$
are points of coexistence between three orientations.  Such an interpretation
of
edges and corners as coexistence lines and points is
important because it means that once we know the details of the ECS,
we can immediately determine into which surface orientations an unstable
surface will decompose.  The consequences of this will be further spelled
out in the next section where we consider the phase separation process in
detail in the two different temperature regimes $T < T_{\rm CR}$ and
$T_{\rm CR} < T < T_{\rm ER}$.

\section{Phase Separation for the case of solid--vapor equilibrium}\label{sec3}

We first look at a few snapshots of equilibrium configurations of a
surface of orientation $[15,5,1]$ in the [111]--RSOS model.  In
Fig.\ \ref{Separated}(a), we see that for $T < T_{\rm CR}$ the configuration
consists of a surface which (apart from a few thermal fluctuations) is
phase--separated into $[100]$, $[010]$, and $[001]$ facets.  In
Fig.\ \ref{Separated}(b), at a temperature
$T_{\rm CR}< T < T_{\rm ps}([15,5,1])$ where $T_{\rm ps}([15,5,1])$
is the phase--separation temperature for this surface, it appears to have
separated into one phase vicinal to the [100] facet and another vicinal to
the [010] facet.  Finally, at a temperature $T > T_{\rm ps}([15,5,1])$,
the surface is stable and does not phase separate.  These
qualitative observations set the stage for our detailed study of the
equilibrium and dynamic aspects of the phase separation problem in the two
regimes,
$T < T_{\rm CR}$ and $T_{\rm CR}< T < T_{\rm ps}([hkl]) \le T_{\rm ER}$, in
which phase separation occurs for an arbitrary surface $[hkl]$.

\subsection{Below $T_{\rm CR}$}\label{sec3a}

In the regime $T < T_{\rm CR}$, all surfaces except the $\{100\}$ facets
are unstable.  Any surface prepared in another orientation
will decompose into a combination of these surfaces.
We can determine which surfaces by
utilizing the mapping of the ECS onto a phase diagram discussed
above.  For example, an arbitrary surface $[hkl]$ with $h$, $k$ and $l$ all
positive is a surface which ``lives" in the three--phase coexistence
region represented by the corner of the ECS where the $[100]$, [010], and
[001] facets meet.  Therefore, such a surface will decompose into a linear
combination of these three surface orientations, with the amount of each
orientation
determined by the requirement that the resulting surface still has
the average orientation $[hkl]$.  (Electron micrographs of the faceting
of the [111] surface of NaCl can be found in Ref.\ \cite{Knoppik}, but see also
the comments in Ref.\ \cite{Heyraud}.)  For the special case of a [hk0]
surface, the surface ``lives" in the two--phase coexistence region represented
by the entire edge between the [100] and [010] facets and therefore
decomposition occurs into just these two surface orientations.

In Ref.\ \cite{OurLongPaper}, we gave
arguments for logarithmically slow growth of the facet sizes
below $T_{\rm CR}$ for the case of the decomposition of a
$[111]$ surface.  There, we also presented strong numerical evidence from
Monte Carlo simulations supporting this claim.  Although the discussion there
was for evaporation--condensation dynamics and within the [111]--RSOS model,
the same arguments should hold outside the RSOS approximation and should also
apply for the case when the dominant mechanism is surface diffusion (in which
case the RSOS model is too restrictive to be used).  The arguments also
generalize to the case of decomposition of any arbitrary surface $[hkl]$.

While the reader is referred to Ref.\ \cite{OurLongPaper} for the
numerical evidence supporting our claim, the basic argument itself is
repeated here for completeness: Consider
a coarsening surface such as that shown in Fig.\ \ref{surface}.  At a time $t$,
the characteristic length scale ({\it i.e., } average facet size) is $L(t)$.
In order for the structure to coarsen further, a step across must propagate
across one of the facets; for definiteness, say a [100] facet.
Since the step consists of, microscopically, pieces of [010] and [001] surface
orientations embedded in the [100] surface orientation,
the step consists of (again, on the microscopic level) edges between facets
of different orientation, and thus it costs an energy per unit length.  Once
the effects of entropy are considered, we find that
there is still a nonzero step free energy ({\it i.e.,} a free
energy per unit length) up to $T = T_{\rm CR}$.
In fact, within the RSOS approximation, this step free
energy as a function of angle can be calculated exactly  (See appendix).  It
is only at $T_{\rm CR}$ that this step free energy goes to zero for a step of a
certain angle (specifically, a $45^\circ$ step).

The coarse--grained differential equation for the growth of $L(t)$ during the
coarsening process can be written as \cite{OurLongPaper}
\begin{equation}
        {dL\over{dt}} = {a(L,T) \over L^m} \ . \label{DifferentialForm}
\end{equation}
Here, $m$ depends on the specific type of dynamics to be considered.  For
example, for evaporation--condensation dynamics, we likely have $m = 2$ or
$3$ (See Sec.\ B).
However, the result below for the asymptotic growth law is independent of $m$.

In the standard case where the kinetic coefficient $a(L,T)$ has no
(or only very weak) dependence on $L$, Eq.\ (\ref{DifferentialForm}) gives
$L(t) \sim t^{1/(m+1)}$.
However, here, the fact that the coarsening involves activation over energy
barriers of height $f_B(T) L$ implies that the kinetic coefficient $a(L,T)$
decreases exponentially with $L$:
\begin{equation}
	a(L,T) = a_0 e^{-f_B(T) L / T} \ .
        \label{KineticCoefficient}
\end{equation}
$f_B(T)$ is a free energy barrier per unit length, and is related to the step
free energy as discussed at the end of the appendix.
At asymptotically long times, Eqs.\
(\ref{DifferentialForm}) and (\ref{KineticCoefficient}) imply
\begin{equation}
        L(t) \sim {T \over{f_B(T)}} \ln(t)
        \ .
        \label{LogGrowth}
\end{equation}

It is worthwhile to consider this result
in the context of the recent work by other groups on this problem.  Monte Carlo
simulations of the coarsening process for $[hk0]$ surfaces, using essentially
the same model as we have discussed above, have
recently been presented by Vlachos, Schmidt, and Aris\cite{Vlachos}.
They consider both the case of the crystal in equilibrium with the vapor
and the case where there is a chemical potential difference between them.
For the equilibrium case, they found generally slow coarsening with an
effective exponent $n_{\rm eff}$ in $L(t) \sim t^{n_{\rm eff}}$ which is small
and decreases with decreasing temperature.  This scenario is consistent with
the behavior found in previous studies of models believed to coarsen
logarithmically\cite{OurLongPaper,GrestAndSrolovitz}.  Since small exponents
are hard to distinguish from a logarithm (and the logarithmic form is expected
only asymptotically), the ability of Vlachos {\it et al.\ }
to fit their numerical data to power law growth with a small exponent should
not surprise us.

We should note that the claim by Vlachos {\it et al.\ } that a logarithmic
form is inconsistent with their data \cite{Vlachos} is intended to mean
only that a logarithm cannot be fit over the entire time regime
\cite{VlachosPrivate}.  In fact, from the inset of their Fig.~2(b), we see
that a logarithm fits quite well (at least as well as a small power
law) over the later time regime ($t > 10^2$).  The data for shorter times are
in the so--called ``facet nucleation regime" \cite{Vlachos}
and can be fitted by neither a logarithm nor a power law.  This regime can be
understood as the characteristic activation time for depositing or desorbing
particles on the initial surface, since the smallest energy barrier of
$4 J_1$ implies a characteristic time on the order of
$t \sim \exp(4 J_1/T) \sim 55$
to successfully deposit the first layer on the surface.  (Here we use the
fact that the energy scale $w_1$ in Ref.\ \cite{Vlachos} is related to ours
by $w_1 = 4 J_1$.)

Other results presented by Vlachos {\it et al.,\ } such as the speeding up
of the growth rate when a chemical potential difference exists between crystal
and vapor, are at least in qualitative agreement with our picture
(See Sec.\ \ref{sec4}).  Thus, we
can say that the simulations performed in Ref.\ \cite{Vlachos}, while
certainly not providing very conclusive support for our claims, are at least
consistent with them.

The continuum approach of Liu and Metiu \cite{Fong}, by contrast, clearly
predicts power--law growth of the facet sizes.  How are we to reconcile
this difference?
Our assertion is that this continuum approach misses a fundamental part
of the problem which proves vital to the dynamics in the discrete
models (and, we believe, although we are somewhat less certain, in the
experimental systems).  In particular, in those cases
where an unstable surface decomposes into surfaces which are smooth
({\it i.e.,} below their roughening transition),
the dynamics of the faceting of surfaces generically involves the
propagation of a step across the facet.  Since such a step has a nonzero
free energy per unit length for $T < T_{\rm CR}$, this leads to
length--scale--dependent barriers and logarithmically--slow growth.

Such a dependence is not
captured by a continuum model which does not recognize that there is a
smallest size, {\it i.e.,\ } the width of a step, determined by the
discreteness of the system.  Rather, in Ref.\ \cite{Fong} the authors introduce
a term into their free energy which imposes an energy cost for rapid changes
in surface orientation.
This captures some of the physics that arises in the microscopic models,
but clearly not all of it.  Furthermore, because the Langevin equations cannot
cope with singularities (and in keeping with the spirit of coarse--graining),
they round the cusps in the surface tension associated with the crystal facets.
This means, in effect, that there is no longer any roughening transition in the
model: all surface orientations are rough!

The idea that the discreteness of a crystal is fundamental in determining
dynamical, and even equilibrium, behavior is certainly not without precedent.
The very existence of a roughening transition and of nucleated dynamics for
crystal growth below this transition is dependent upon including terms in the
Hamiltonian (and the resulting dynamical equations) which explicitly model
the discreteness of the system
\cite{ChuiWeeks,WeeksGilmer,Nozieres,VanBeijerenNolden}.
As far as we know, there is
no known prescription for determining {\it a priori} whether such discreteness
will be relevant or irrelevant in a renormalization group sense.
In the absence of such a prescription, purely continuum approaches to the
problem (even if they mimic some effects produced by the discreteness,
{\it e.g., } by using anisotropic surface tensions and energy penalties for
changes in surface orientation) should, we believe,
be used with a certain degree of caution.

In Sec.~IV, we will return to the case of coarsening below $T_{\rm CR}$,
but in the
case where the crystal is slightly out of equilibrium with the vapor.  There,
we will find that the difference in chemical potential between vapor and
crystal
sets a maximum size to the free energy barrier, so the problem becomes one
of nucleation and there is a return to power--law behavior at late enough
times.
However, first we will consider the coarsening process for the case of
crystal--vapor equilibrium but in the temperature regime above $T_{\rm CR}$.

\subsection{Above $T_{\rm CR}$}\label{sec3b}

When we consider the diagram in Fig.\ \ref{AllowedOrientations}
showing the stable orientations at a temperature $T_{\rm CR} < T < T_{\rm ER}$,
three questions immediately present
themselves: (1) How does one determine the location of the boundaries
between the stable and unstable surface
orientations?  (2) For a surface in the unstable region, what are the
orientations of the stable surfaces it breaks up into?  (3) What are the
kinetics of this process?

\subsubsection{Equilibrium aspects}\label{sec3b1}

We find it most natural to begin with a discussion of the second question
raised above.  The behavior of an unstable surface
depends on the details of the ECS.  In
Figs.~\ref{ridge} and \ref{conicalpoint}, we consider two possible scenarios
\cite{OnlyTwo}.

In Fig.~\ref{ridge}(a), for what we dub the ``ridge scenario", the sharp edges
(``ridges") between the facets continue into the curved region of the ECS.  As
one moves along such a ridge into that region, the angle between the unit
vectors normal to the two surfaces that meet at that point on the ridge
decreases from $90^\circ$ at the facets to $0^\circ$ at a
second--order point at which the ridge terminates. Since at any point along
this
ridge two surface orientations meet, the mapping of the ECS onto
a phase diagram tells us that this means there is coexistence between two
surface orientations at each point.
Without loss of generality, we consider the case of a surface $[hkl]$ with
$h \ge k \ge l > 0$.  Fig.~\ref{ridge}(b), showing part of
the diagram of stable surface orientations, demonstrates how such a surface
prepared at an unstable orientation breaks up into two surfaces.  Note that,
by symmetry, the two coexisting surfaces are symmetric about the line $h=k$.
Also, a linear combination of these two surface orientations must add up to
give an average surface orientation of $[hkl]$.  These two conditions,
along with the third condition specifying those surfaces which are on the
curve of marginally stable surfaces, completely specify the two surfaces
which the surface $[hkl]$ decomposes into.

In Fig.~\ref{conicalpoint}(a), which we dub the ``conical point scenario", the
sharp edge terminates at the point where it meets the facet boundaries.  (This
can be considered as a limiting case of taking the length of the ridges into
the curved regions to zero.)  In this case, all the unstable surfaces must
``live" in the point where the edge meets the facet boundaries.  Furthermore,
all the marginally--stable orientations along the boundary in
Fig.~\ref{conicalpoint}(b) must meet at this one point on the ECS.
The slope of the ECS at this point depends on the angle from which it is
entered.  Therefore, in the rounded region of the ECS, the point has the
symmetry of the tip of a cone, hence the name ``conical point."
Also note that the ECS has a jump in
slope as one goes from the rounded region of the ECS, through the
conical point, and onto one of the facets (or, alternately, along
the sharp edge between the two facets) \cite{SurfaceTension}.

Naively, one might expect that the ridge scenario would occur
in general, as the conical point scenario seems to require that
there be special symmetry about the conical point, which results in the meeting
of all the different orientations at this single point on the
ECS \cite{PottsModel}.  Such a point is very much analogous to the
zero field (${\bf H} = {\bf 0}$) point in the low temperature phase of a
three--dimensional XY model, at which an entire one--parameter family (circle)
of magnetizations coexist.
However, in that model such degeneracy is clearly the result of the symmetry
with respect to spin orientation $\theta$ which occurs for ${\bf H} = {\bf 0}$
in the original Hamiltonian.  In the present problem, there is no such obvious
symmetry and thus such a point would not be expected unless the
terms breaking this symmetry become irrelevant under renormalization of the
original Hamiltonian.

To rigorously determine which scenario occurs in one specific model, we have
recently studied \cite{SixVertex} the [110]--RSOS model for the surface of an
FCC crystal with nearest--neighbor attractive and next--nearest
neighbor repulsive interactions\cite{JayaprakashSaam}.  In this model,
$T_{\rm CR} = 0$, i.e., the corners round for any $T > 0$.  However,
a sharp edge separating the $[111]$ and $[11\overline{1}]$ facets persists
up to a nonzero temperature $T_{\rm ER}$.
The advantage of this model is that it can be mapped onto the six--vertex
model\cite{JayaprakashSaam,Unphysical} which is exactly solved
\cite{SixVertexSolution,SixVertexReview}.
The solution of the six--vertex model, which is given in terms of integral
equations, can be investigated numerically and, in certain limits,
analytically \cite{SixVertex}.  We find that in this model, the conical point
scenario occurs, as was conjectured by Jayaprakash and
Saam\cite{JayaprakashSaam} at the time when they first considered this
RSOS model \cite{SixVertexScenerio}.

We also find an exact expression for the curves
separating the stable and unstable surface orientations,
as a function of temperature. (These curves are analogous to the boundaries
between the shaded and unshaded regions for $T_{CR} < T < T_{ER}$ shown in
Fig.~\ref{AllowedOrientations} for the simple cubic case).
As one moves along the ECS away from the conical point into the curved region,
the crystal shape has the following nonanalytic behavior:
\begin{equation}
	F(h) = F_0 + F_1 \delta h + F_2 \delta h^{5/3}+{\cal O}(\delta h^{7/3})
        \ ,
        \label{scaling}
\end{equation}
where $\delta h$ is the distance from the conical point along the
$h$--direction and for simplicity we have shown the dependence on only one of
the two position variables $(h,v)$.  Here, the crystal shape has been oriented
such that $F(h,v)$ is constant along the first--order edge between facets.
The second term then corresponds to the jump in slope at the conical point.
The
third term shows, however, that there is more fundamental nonanalytic behavior,
namely that an analytic expansion about this point does not exist, even if one
requires that the expansion only be valid in the
curved region of the ECS / free energy surface.
(Analogous nonanalyticities appear in the surface tension $F(x,y)$
\cite{SixVertex}.)  Nonanalyticities in the free energy
[beyond the expected discontinuities in the derivatives of $F(h,v)$] at
first-order coexistence boundaries have been discussed previously\cite{Binder}
and are thought to be a rather generic feature in phase diagrams.
They are absent within
mean field theory, where first--order transitions are associated simply with
crossings of the local minima in the free energy surface.
It is only through the effects of renormalization that such nonanalytic
behavior can appear.  In the Ising model, the nonanalyticities in the free
energy $F(H)$ at zero field (H = 0) for $T < T_C$ are weak essential
singularities.  For vector spin models, as here, power--law singularities are
observed.  These divergences can be explained using spin wave theory
\cite{Binder}.

Clearly, the question which must be asked is how generally a conical point
occurs, as opposed to having a ridge extend into the rounded region of the ECS.
We are not able to answer this question rigorously.  However, we believe that
the evidence from the six--vertex model suggests the conical point will be
a general feature of equilibrium crystal shapes at the point where a sharp
edge between facets meets a rounded region of the ECS.  This belief
is based upon the following observation:  In the six--vertex model, the
appearance of the conical point, having such a high degree of symmetry, does
not seem to be due to any special symmetry which already exists in
the original Hamiltonian.  Rather, the symmetry seems to be generated
spontaneously under renormalization (and this occurs for {\it all}\
temperatures and values of the interaction parameters that put us in the
``low--temperature ferroelectric regime" of the model).  Therefore,
we expect such conical
points to be a rather generic feature of equilibrium crystal shapes,
probably occurring for simple cubic as well as FCC models and outside the RSOS
approximation.  On the other hand, Neergaard and Den~Nijs have recently
presented an argument that the conical point, or at least the specific
scaling behavior at this point, may be a rather special feature of the
six--vertex model \cite{NeergaardDenNijs}.  In light of these conflicting
opinions, we must say that the generality of this feature remains an open
question.

It is interesting to note that the conical point in the crystal shape and the
associated coexistence region in the space of surface orientations is precisely
the converse of a cusp in the surface tension and the associated facet on
the crystal shape \cite{VanBeijerenPrivate}.  In this latter case, the conical
point (cusp) in $F(x,y)$ corresponds to a flat facet in $F(h,v)$.  The
analogue of the one--parameter family of orientations coexisting for the
former case is the one--parameter family of points on the crystal shape
(namely, those on the boundary of the facet) which all correspond to the cusp
point in $F(x,y)$.  Also, the sharp edge, or ridge, between the facets
in $F(h,v)$ is analogous to the ``groove" in the surface tension $F(x,y)$
which exists between the two coexistence regions associated with the two
conical points \cite{SixVertex}.

We are now ready to discuss the first
question posed at the beginning of Section~\ref{sec3b},
namely the determination of the boundary between stable and unstable
orientations as a function of temperature.  As noted above, this boundary
can be determined exactly for the FCC case, within the [110]--RSOS
approximation, because of the mapping onto the exactly--solved
six--vertex model.  For the
simple cubic case, we do not have an exact solution even within the [111]-RSOS
approximation.  However, we can calculate the ``opening angle" $\theta_c(T)$
for this curve.  [In Fig.\ \ref{conicalpoint}(b), this is the opening angle
between the bold solid line and the octant boundary.]
This angle is also half the opening angle $2 \theta_c(T)$ between the two
facet boundaries on the ECS at the conical point.
Because calculation of $\theta_c(T)$ involves surfaces which are vicinal
to the facet orientations ({\it i.e.,} the steps across the facets are widely
spaced so that step--step interactions are not important), it can be determined
from an exact calculation of the free energy of an isolated step within the
[111]-RSOS approximation.
The calculation of the step free energy, $f_s(T,\theta)$ and the angle
$\theta_c(T)$ are
discussed in the appendix.  An alternate way to view $\theta_c(T)$ is that
its inverse function, $T_{\rm ps}(\theta)$, gives the ``phase separation
temperature"
for a vicinal surface ({\it i.e., } that temperature below which the
surface is unstable) as a function of the angle of the steps across the
surface (and thus its orientation).  A plot of $T_{\rm ps}(\theta)$ is given in
Fig.\ \ref{Angle}.

We have carried out some simulations of the breakup of various surfaces $[hkl]$
with $h \ge k \ge l > 0$ at various temperatures.  Those surfaces studied which
are in the limit $l \ll h$ ($[15,5,1]$, $[22,7,1]$, and $[30,29,1]$ surfaces)
appear to decompose into two phases, one vicinal to the $[100]$ surface and
the other vicinal to the $[010]$ surface
[see, {\it e.g.}, Fig.\ \ref{Separated}(b)].  When the system is quenched from
high temperatures, thin stripes of these two vicinal phases form and then
coarsen over time until only one stripe of each phase remains in the system.
We have measured the approximate angles of the steps across these vicinal
phases and find that they are in quite
good agreement with our formula for $\theta_c(T)$ [see Fig.\ \ref{Angle}].

Note that the fact that the surface appears to decompose
into only two surface orientations, rather than an entire one--parameter
family, seems to be in contradiction to the conical point scenario we have
argued is likely to apply.  However, three important points should be stated.
The first is that it is, in fact, hard to tell if there are really only two
orientations present.  What we can say with confidence is
that there seems to be a breakup into a ``phase"
with orientations near $[100]$ and a ``phase" with orientations near $[010]$,
and that both these phases consist of vicinal surfaces with steps having
approximately the
angle $\theta_c(T)$ across them.  This is not really in contradiction with
what we expect: In the conical point scenario, the decomposition will consist
of a weighted distribution of the surfaces in
the one--parameter family of marginally stable orientations.
For decomposition of a surface close to being a $[hk0]$
surface, the decomposition must primarily consist of those surfaces
vicinal to the $[100]$ and $[010]$ surfaces in order to get the correct
average surface orientation.  (For a $[hk0]$ surface itself, the
decomposition will simply be into the [100] and [010] surface orientations.)

Our second point is that when we then carry out simulations of surfaces which
are not as close to being an $[hk0]$ surface (and at temperatures nearer to
$T_{\rm CR}$ in order that we see any phase separation at all), then the
equilibrium
configurations do not appear to be so simple.  Fig.\ \ref{NotVicinal} shows
an equilibrium configuration of a $[13,13,4]$ surface at $T = 8 J_2$.
At $T = 9 J_2$, the configurations are even more complicated and it is in fact
hard to tell if the surface has phase--separated at all.  Unfortunately,
in studying these configurations, we have found it
difficult to quantify the surface phase separation since one
would have to measure the surface orientation coarse--grained over some region,
because (unlike, {\it e.g.}, in the case of an XY model) the order parameter
is not a continuous variable at the microscopic level: each rhombic tile
represents either a [100], [010], or [001] orientation.
The correct scale over which to
coarse--grain is not obvious, and of course, the larger this scale, the larger
the system size should be in order to avoid finite--size constraints.

Our final point is that
the periodic boundary conditions put some constraints on the system,
specifically, that the stripes of each ``phase" in Fig.\ \ref{Separated}(b)
must be horizontal.  This may be influencing the results considerably.
Therefore, one might need to go to much larger system sizes
(which is not computationally feasible at present)
in order to start truly seeing the other available surface
orientations.  Another approach would be to try to choose alternative
boundary conditions (such as free or helical).
However, implementing such boundary conditions and forming
the initial surface configuration is a nontrivial task in this model.

In the final analysis, we would have to say that the simulations currently do
not provide conclusive evidence either for or against the conical point
scenario in this model.

\subsubsection{Kinetics}\label{sec3b2}

To study the kinetics of the phase ordering process, we turn to the continuum
formalism used by Liu and Metiu \cite{Fong}.  For the case of the
evaporation--condensation mechanism, and neglecting noise since temperature
is believed to be an irrelevant variable, they derived the
Langevin equation
\begin{equation}
	{\partial{\bbox{\psi}}({\bf r},t)\over\partial t}
		= - A \biggl \lbrack a \nabla^4
		{\bbox{\psi}}({\bf r},t) - \nabla \bigl ( \nabla \cdot
		{\partial F_0(x,y)\over\partial{\bbox{\psi}}}
		\bigr ) \biggr \rbrack
        \ ,
        \label{langevin}
\end{equation}
where ${\bf r} \equiv (h,v)$ and ${\bbox{\psi}} \equiv (x,y)$.
(Here, we have used the constraint
$\nabla \times {\bbox{\psi}} = 0$ to rewrite the first term
on the right hand side \cite{SiegertPlischke}.  This constraint follows from
the fact that ${\bbox{\psi}}$ is the gradient of the surface.)
If surface diffusion is the dominant mechanism, then there is an additional
$-\nabla^2$ in front on the right hand side.  $F_0(x,y)$ is some
suitably coarse--grained projected surface tension, not the full
thermodynamic projected surface
tension $F(x,y)$.  In particular, $F_0(x,y)$ is not necessarily convex since,
in the case when the $(x,y)$ orientation is unstable, it reflects the free
energy of metastable configurations having orientation $(x,y)$ rather than the
thermodynamic configurations once phase separation has occurred
(See, {\it e.g., } Ref.\ \cite{Langer}).

Because the order parameter ${\bbox{\psi}}$ is conserved, the stable stationary
states ({\it i.e.,\ } those orientations into which a surface decomposes) are
not simply given by the minima of $F_0(x,y)$.  Rather they are given by
those values of ${\bbox{\psi}}$ that share
a common support tangent on $F_0(x,y)$ (See, {\it e.g., }
Fig.\ 3 in Ref.\ \cite{Fong}).

Note that the structure of Eq.\ (\ref{langevin}) is {\it very} similar to that
for the traditional phase separation process, {\it i.e., } the so-called
Cahn--Hilliard Equation (See \cite{Bray}).  The one difference is the reversal
of the ordering of the two vector operators in the second term on
the right hand side:  Here we have the gradient of a divergence (and also
the constraint that $\nabla \times {\bbox{\psi}} = 0$),
while in the traditional case one has the divergence of a gradient
\cite{SiegertPlischke}.  The former
case causes a coupling of the different components of the order parameter, and
of the spatial and order--parameter degrees of freedom, while
the latter case leaves them uncoupled.  (In fact, the former case only makes
sense if $n=d$ where $d$ is the spatial dimension of the interface and $n$ is
the number of degrees of freedom of the order parameter.)

Before considering the case of most interest to us at present (where
phase separation is expected to occur into a continuous family of
orientations),
we will first discuss the ``simpler" case of separation into a finite number
of orientations.  This has been the subject of much very recent work.
In Ref.~\cite{Fong}, the authors assumed a form for the surface tension which
results
in a few ({\it e.g., } three) stable stationary states in $F_0(x,y)$.  On the
basis of simulation results and power--counting, they argued that the growth of
the domain size likely satisfies $L(t) \sim t^{1/4}$.  However, in the case of
traditional phase separation ({\it i.e., } governed by the Cahn--Hilliard
Equation) into a finite number of phases, such power--counting is known
to fail.  Because of the sharpness of domain walls, one finds instead
$L(t) \sim t^{1/3}$.  The most general arguments for these results are given
by Bray \cite{Bray}, who argues that the growth law in the case of conserved
dynamics can be determined from the energy cost of a domain wall.  If
the energy cost goes like $E(L) \sim L^y$ then $L(t) \sim t^{1/(2+d-y)}$.
For the case of sharp domain walls, {\it e.g.,\ } in an Ising model,
$y=d-1$; whereas in the case of domain walls which themselves
are spread out over a width $L$, as in the isotropic vector spin models,
$y=d-2$.

Do Bray's arguments continue to hold for Eq.\ (\ref{langevin}), or does the
reversal in the order of the vector operators result in different behavior?
It appears that, in fact, the renormalization flow equations change such that
simple connection between $y$ and the growth law exponent no longer necessarily
follows \cite{SiegertPlischke}.  Indeed,
in addition to Liu and Metiu \cite{Fong}, a few other groups have
recently simulated equations similar to Eq.~(\ref{langevin}) and have
found results most compatible with $L(t) \sim t^{1/4}$
\cite{SiegertPlischke,KrugDobbs}.  Thus it seems possible that the order of
the vector operators does change the kinetics of the phase separation
process.  On the other hand, given the history of low measured exponents
(often around 1/4) in Ising systems where eventually the exponent was shown
to cross over to 1/3 \cite{History}, we believe that, in the absence of further
analytical understanding, both $L(t) \sim t^{1/4}$ and $L(t) \sim t^{1/3}$ (or
even some intermediate exponent!) should still be considered as viable
candidates for the asymptotic behavior.

Finally, we note that we have also published some of our own numerical evidence
bearing on this question.  In Ref.\ \cite{OurLongPaper}, we looked
at the behavior of the characteristic length scale for the coarsening of a
[111] surface cooled slowly {\it at a constant rate} $\Gamma$.
Assuming the ``underlying growth law" in the absence of diverging barriers
would be $L(t) \sim t^{1/3}$, we argued that
the behavior under asymptotically slow cooling would be
$L(\Gamma,T=0) \sim \Gamma^{-1/4}$.  It was noted that this expectation was in
reasonable agreement with our numerical results.  Here, we add two comments:
(1) If the ``underlying growth law" is instead $L(t) \sim t^{1/4}$, then the
corresponding behavior at a constant cooling rate is
$L(\Gamma,T=0) \sim \Gamma^{-1/5}$
(See endnote 68 of Ref.\ \cite{OurLongPaper}).
(2) The simulations performed there, and further simulations carried out on
systems as large as $240^2$ \cite{Unpublished}, cannot conclusively
distinguish these two possibilities, with the best estimate of the exponent
of $\Gamma$ being $-0.23$, {\it i.e.,} about halfway between the
two expectations.

Of course, the considerations of the previous three paragraphs apply to the
case
where the surface tension is such that phase separation occurs into a
finite number of different orientations separated by sharp domain walls.  This
will be relevant to our discussions in Sec.\ \ref{sec4}.  In the present
case, however, our $F_0(x,y)$ is expected to be of such a form that
we have phase ordering into a continuous one--parameter
family of orientations.  The numerical work of Siegert and
Plischke \cite{SiegertPlischke} suggests that in this case, as in the standard
case when the vector operators are reversed ({\it e.g., }
an isotropic XY or Heisenberg model), the exponent for the growth law
is $1/4$. (Note however that, technically speaking, the case $n = d = 2$
is excluded from Bray and Rutenberg's theory of phase ordering in the
Cahn--Hilliard equation because of long--range correlations between the
topological defects\cite{Bray}.  Simulations of this case are nonetheless
consistent with $L(t) \sim t^{1/4}$, but perhaps with logarithmic corrections
and deviations from scaling \cite{MondelloGoldenfeld}.)

It is worth emphasizing that
the whole decomposition process here is quite different from the case where
separation occurs into a finite number of orientations:  Here, there are really
no well--defined
domains or ``facets" formed at all; rather, it is the length scale over which
the surface bends that will increase over time (which can be measured most
easily from a correlation function or its Fourier transform).  Thus terms
such as ``faceting" and ``phase separation" seem somewhat inappropriate.

\section{Crystal growth or etching}\label{sec4}

We now consider the case in which the crystal is slightly out of
equilibrium with its vapor, that is, the case of slow growth or etching.
In fact, it is known that, far out of equilibrium, the nonequilibrium effects
can drive the formation of facets even when such faceting would not occur in
equilibrium \cite{SiegertPlischke,KrugDobbs}.  Here, however, we will assume
that we are close enough to equilibrium that the stability of a surface is
still determined from equilibrium considerations, and will study the effect
of a chemical potential difference $\Delta\mu$ between solid and vapor phases
\cite{Vlachos} on the dynamics of the phase separation.  In the magnetic
language of the 3-D Ising ferromagnet, such a situation can be thought of as
the application of a uniform magnetic field in the model.

First we consider the temperature regime $T < T_{\rm CR}$ and,
for definiteness, we discuss the case
of crystal growth ($\Delta \mu < 0$). Then the free energy cost
associated with adding surface atoms to produce a step
across a facet can be written approximately as follows:
\begin{equation}
	\Delta F \approx f_p(T,\theta) L - |\Delta\mu| {L^2\over{2}}
	\tan(\theta) \ .
	\label{DeltaF}
\end{equation}
Here $f_p(T,\theta)$ and $L$ are the projected step free energy and the
projected length of the step, respectively (see appendix).  The projection is
onto one of the lattice axes, and $\theta$ is the angle of the step relative to
that axis.  [Eq.\ (\ref{DeltaF}) is valid when the linear size of the facet
itself is larger than $L$.]  The structure of the problem is the same as that
of determining the growth rate (due to nucleation) of a growing crystal surface
at a temperature below its roughening transition\cite{WeeksGilmer}: Assuming
that $|\Delta \mu|$ is not too large, then for small $L$ the first term in
the expression dominates and $\Delta F$ increases linearly with $L$.  However,
for large enough $L$, the second term dominates and $\Delta F$ is a decreasing
function of $L$.

To find the smallest possible barrier associated with propagating a step across
an infinite facet, we want to find the length of step $L_0$ and the angle
$\theta_0$ such that the free energy cost is maximized as
a function of $L$ and minimized as a function if $\theta$. This we do by
setting $\partial (\Delta F)/\partial L
= \partial (\Delta F)/\partial\theta = 0$.
We find
\begin{equation}
	L_0 \approx {f_p(T,\theta_0) \over{|\Delta \mu| \tan(\theta_0)}} \ ,
	\label{L0}
\end{equation}
with an implicit expression for $\theta_0$ given by
\begin{equation}
	{\partial f_p(T,\theta)\over{\partial[\tan(\theta)]}}
	\bigg\vert_{\theta = \theta_0}  =
	{f_p(T,\theta_0) \over{2 \tan(\theta_0)}}
	\ .
	\label{ThetaSpecial}
\end{equation}
However, this latter expression always has the solution $\tan(\theta_0) = 1$,
independent of $\Delta \mu$ or $T$, so a step of $45^\circ$ always gives the
smallest barrier.  Thus, (\ref{L0}) becomes
\begin{equation}
        L_0 \approx f_p(T,\theta=\pi/4) / |\Delta \mu| \ ,
\end{equation}
and the free energy barrier is
\begin{equation}
	\Delta F_{\rm max} \approx {f_p^2(T,\theta=\pi/4) \over{2 |\Delta \mu|
	}} \ .
	\label{Barrier}
\end{equation}

Eq.\ (\ref{Barrier}) gives the maximum size of the barriers for a coarsening
surface.  The barriers to coarsening will grow, and thus the coarsening of the
surface will look roughly logarithmic, out to a time of order
\begin{equation}
	t_c \approx \exp[f_p^2(T,\theta=\pi/4)/(2 |\Delta \mu | T)]
	\label{CharacteristicTime}
\end{equation}
when the characteristic facet size $L$ is of order $L_0$.  At times longer
than this, the barrier height to coarsening remains saturated at $\Delta F_{\rm
max}$, independent of $L$, and the growth will thus proceed with a power law
\begin{equation}
	L(t) \sim (\gamma t)^n \ ,
\end{equation}
with $\gamma \approx \exp[-f_p^2(T,\theta=\pi/4)/(2|\Delta\mu | T)]$.
Here, $n$ is the exponent for growth associated with Eq.\ (\ref{langevin}) in
the case where the surface breaks up into a finite number of orientations
(thus, either $n = 1/3$ or $1/4$, as discussed in Section\ \ref{sec3b2}).

The form of $\gamma$ is closely analogous to that for the rate of
growth of a defect--free crystal surface below the roughening transition
\cite{WeeksGilmer}.  The one important difference, however, is that in this
latter case, no driving force exists for the growth of the crystal in the
limit $\Delta\mu = 0$.  In the case of the phase separation problem, however,
while there is still no driving force {\it for growth of the crystal} in the
absence of a chemical potential difference, there {\it is} a driving force
{\it for phase separation of the surface}.  Thus, the characteristic length
scale grows with time even when $\Delta\mu = 0$, but the process
occurs logarithmically--slowly in this case.

We have performed Monte Carlo simulations of the [111]--RSOS model in order to
test certain aspects of our above arguments.  First, in order to study the
tenet of our argument involving the behavior of the free energy barrier as
a function of the size of the facet [Eq.\ (\ref{DeltaF})],
we consider the process of removing
one face of a ``cubical projection" from a crystal surface, as illustrated in
Fig.\ \ref{CubicalProjection}.  Fig.\ \ref{ShrinkingProjection} is an
Arrhenius plot of the time such a process takes as a function of the linear
size $L$ of the projection for the case $\Delta\mu = 0$.  (Cf. Fig.~6
of \cite{OurLongPaper} where a similar plot was presented for the case of
fully three-dimensional coarsening.  Also see that reference for further
details about the simulation methods.)  We see that, as is our expectation, the
slope on the Arrhenius plot, and thus the free energy barrier, is an increasing
function of the size of the projection.  In fact, the expected free energy
barrier in the limit $T \to 0$ can easily be calculated \cite{MyThesis} to be
\begin{equation}
	F_B(L,T=0)=\left\{
	\begin{array}{ll}
	12 J_2\ , & L=2\ .\\
	4 J_2 (L+2)\ , & L > 2\ .
\end{array}\right.
\label{BarrierNo}
\end{equation}
Fits to the form $t = t_0(L) e^{F_B(L,T=0)/T}$, with $t_0(L)$
as a free parameter, are also shown in the figure and are generally very good
for the low--temperature data.  Only for $L = 8$ are systematic deviations from
the fits still evident at the lowest temperatures, with the slope on the plot
being {\it greater} than that predicted.  This deviation is due to our
approximation of $F_B(L,T)$ by $F_B(L,T=0)$ (See Sec. II of
Ref.\ \cite{OurLongPaper}).

In Fig.\ \ref{ShrinkingProjectionMu}, we present an Arrhenius plot for the
time to shrink a cubical projection of size $L$ for $\Delta\mu/J_2 = 2$.
(Here, the sign of $\Delta\mu$ is chosen to be positive, {\it i.e.,\ } to favor
shrinking of the projection).  The contrast to Fig.\ \ref{ShrinkingProjection}
is apparent:  For small $L$, the free energy barrier is
an increasing function of $L$.  However, for $L \ge 6$, the barrier appears
to saturate.  The expected barrier for each $L$ can, in fact, be calculated by
considering the highest energy state the system must pass through in removing
a face on the projection.  This gives
\begin{equation}
	F_B(L,T=0)=\left\{
	\begin{array}{ll}
	12 J_2 - \Delta\mu \ , & L=2\ .\\
	4 J_2 (L+2) - \Delta\mu (L+1)\ , & 2 < L \le 6\ .\\
	18 J_2\ , & L \ge 6\ .
\end{array}\right.
\label{BarrierMu}
\end{equation}
(Here, the saturation of the barrier at $L = 6$ is for the specific case of
$\Delta\mu/J_2 = 2$.)
Fits to the form $t = t_0(L) e^{F_B(L,T=0)/T}$, with $t_0(L)$ as a free
parameter, are also shown in the figure and are generally very good for the
low--temperature data.  (The one fit which shows a bit of deviation is for
$L = 6$, probably because of the presence of a large degeneracy of barrier
states with very similar barrier heights.)
Also, the value of $18 J_2$ which we get for
the saturated barrier height is in quite good agreement with the estimate of
$16 J_2$ provided by Eq.\ (\ref{Barrier}) in the limit $T \to 0$ [which can be
worked out using Eq.\ (\ref{ThetaSpecial}) and formulas in the appendix].

The preceding study tested our assertions about the barrier heights to
coarsening for one very artificial kind of configuration.
In order to test more directly the effect of a nonzero $\Delta\mu$ on the
coarsening process, we present in
Fig.\ \ref{Coarsening} full--scale simulations of the coarsening of a [111]
surface, which is initially in a random ($T = \infty$) configuration and
is quenched to $T = 2 J_2$.  The figure shows the growth of the characteristic
length scale $L(t)$ over time for various values of
$|\Delta\mu|/J_2$.  We see that, in its qualitative aspects, the Monte Carlo
data is in agreement with our expectations:  First, we consider the
case $\Delta\mu = 0$.  The growth,
although not yet quite logarithmic as would be expected at asymptotically long
times, is slower than power law; the effective exponent in
$L(t) \sim t^{n_{\rm eff}}$ is $n_{\rm eff} \approx 0.07$ at the latest times
and is continuing to slowly decrease.

As the chemical potential
difference between solid and vapor is increased from zero, the growth becomes
faster.  For $|\Delta\mu|/J_2 = 0.5$, the effective exponent is virtually
constant
over the six decades of time shown.  For the larger $|\Delta\mu|$, the increase
in the effective exponent with time is quite gradual.
However, the most marked bending of the data does appear to occur at times
whose
trend with $|\Delta\mu|$ is in qualitative agreement with (but that seem to be
quantitatively a bit later than)
the expectations of Eq.\ (\ref{CharacteristicTime}).
For $|\Delta\mu|/J_2 = 1.2$ and $2.0$, there is a regime where the effective
power law approaches (although remains a bit less than) $1/4$.

Also visible for $|\Delta\mu|/J_2 = 2.0$, and just barely for
$|\Delta\mu|/J_2 = 1.2$, is a rather sharp leveling off in the growth of $L(t)$
at late times.
Studying the configurations at these late times, we find that the system
appears to have reached a steady--state in which at any time, there
are structures such as ledges, etc. because the crystal is in the process of
growing (or shrinking), due to the nonzero value of $\Delta\mu$.
The fact that such growth is in progress puts a maximum limit
on the average facet size.  [Our measure of $L(t)$ as being inversely
proportional to the total length of boundary between the different facet
orientations may also be underestimating somewhat the true value of the
characteristic length scale.  See Sec.\ IV.C of Ref.\ \cite{OurLongPaper}.]
A more detailed
study of correlation functions might be useful in further characterizing this
behavior ({\it e.g., } one might even find a breakdown of scaling).
Clearly, these are effects which occur in a regime beyond which our assumption
of being very near equilibrium is valid.

In closing this section,
we briefly discuss the likely effects of a nonzero chemical
potential difference in the temperature regime $T_{\rm CR} < T < T_{\rm ER}$.
Here, the coarsening is already expected to be a power law for $\Delta \mu =
0$.
A nonzero $\Delta \mu$ seems unlikely to do more than alter the prefactor
of the power law.  One important question, however, is whether it might provide
a relevant perturbation which will break the symmetry associated with
the conical point and thus allow the selection of
two particular orientations for the surface to separate into.  In the absence
of a renormalization group picture for the conical point on the ECS,
we do not know how to determine if this perturbation is relevant, and hence
this question remains unanswered.

\section{Summary and conclusions}\label{sec5}

In this paper, we have considered the ``phase separation" of crystal
surfaces within an Ising lattice gas model for materials with the sodium
chloride structure.  We found that, depending
on the temperature and growth conditions, a number of interesting behaviors
can be observed.

For a crystal in equilibrium with its vapor,
we argued that the coarsening of the surface structure
will be logarithmic in time for temperatures below the corner rounding
transition temperature $T_{\rm CR}$, where an arbitrary surface $[hkl]$
(with $h$, $k$, $l > 0$) undergoes
three--phase separation into the smooth $[100]$, $[010]$, and $[001]$ facets.
Such sluggish dynamics occur because the coarsening
process involves the creation of steps across the smooth facets and these steps
have a nonzero free energy per unit length below $T_{\rm CR}$.  Since the
discreteness of the system plays a fundamental role here,
recently proposed continuum models for the dynamics of
phase separation of surfaces fail to capture this effect.

At all temperatures between $T_{\rm CR}$ and the edge rounding temperature
$T_{\rm ER}$,
some surface orientations remain unstable.  It appears likely, based on the
exact solution of a related model, that (because of a spontaneously--generated
symmetry) such surfaces will generically undergo
decomposition into an entire one--parameter family of surfaces.  This
process, analogous to phase ordering in an XY (or Heisenberg) model, should
obey $L(t) \sim t^{1/4}$.  Only if effects not considered (such as elastic
interactions or surface melting) break the spontaneously--generated symmetry,
does the possibility exist for simple two--phase separation.  In that case, the
growth law will be $L(t) \sim t^{n}$, with either $n = 1/3$ or $1/4$.

Such an algebraic growth law should also hold at late enough times {\it below}
$T_{\rm CR}$ for the case of {\it driven} surfaces,
{\it i.e.,\ } where the crystal and vapor are not in equilibrium
(crystal growth, or etching), with the crossover time from very slow
(approximately logarithmic) to power--law growth diverging rapidly as the
chemical potential difference between crystal and vapor decreases to zero.
We noted the close analogy in this case to the growth of a crystal
facet by nucleation below the roughening transition.

In closing, we should mention some effects which have been neglected in our
microscopic model and could conceivably be important in real materials.
First, there are impurities and defects which may act to
speed up the phase separation of surfaces, just as they do for the
growth of crystal surfaces below the roughening transition
\cite{WeeksGilmer,Gilmer}.  This could be
particularly important given that impurities often tend to segregate to the
surface, so that even reasonably pure samples can have fairly dirty surfaces.
Second, there are also elastic effects, which have been discussed in recent
work
\cite{Stewart,Fong,Alerhand,Marchenko} as possibly slowing
down the faceting process and even stabilizing the system at a
maximum facet size.  There is some recent experimental evidence which
seems to support this scenario\cite{Phaneuf1}.
Finally, there is the possibility of surface melting
\cite{VanBeijerenNolden,VanBeijerenPrivate} and surface
reconstructions\cite{Yoon} which can result in a more complicated ECS.
Clearly these issues must all be addressed in order to
obtain a realistic picture of the thermal faceting in real materials.

\section*{Acknowledgements}

We would like to thank Alan Bray, Mark Holzer, Fong Liu, Michael Plischke,
Jim Sethna, Martin Siegert, Henk van Beijeren, and
Michael Wortis for helpful discussions.
This work was supported by the NSERC of Canada.

\appendix
\section*{Calculation of the step free energy and related
quantities}\label{appendix}

Here, we calculate the step free energy, $f_s(T,\theta)$,
as a function of temperature $T$ and step angle $\theta$ (measured with
respect to one of the lattice axes) within the [111]--RSOS approximation.
We then use this result to discuss features of the phase separation problem
in the temperature regimes both below and above $T_{\rm CR}$.
The reader is also referred to Ref.\ \cite{ShiWortis} where a calculation
of the step free energy was performed from a somewhat different, but
related, perspective.

It is in general quite difficult to compute the step free energy in a
so-called ``canonical ensemble," where we restrict the step to have a certain
angle.  The easier route is to compute the step free energy in the ``grand
canonical ensemble", where we do not constrain the step angle, but instead
apply a ``field" $\tilde h$ which couples to the angle of the step and thus
allows us to control the average step angle \cite{Holzer}.
To obtain $f_s(T,\theta)$ from $f_s(T,{\tilde h})$
we then need only determine this average step angle $\theta$ as a function
of $\tilde h$.  Then, $f_s(T,\theta)$ is related to $f_s(T,\tilde h)$
by a Legendre transform, which has an intuitive physical interpretation.

In the [111]--RSOS approximation, a step across, say, a [100] surface consists
simply of a one--dimensional series of plaquettes of either the [010] or
[001] orientation [see, {\it e.g.}, Fig.\ \ref{Separated}(b)].
If we apply a ``field" ${\tilde h}$ in order to favor one orientation
over another, then the Hamiltonian for the step is equivalent to
that for a one--dimensional Ising model in a magnetic field.  Within this
grand canonical ensemble, there are several routes to calculate
the step free energy.  If one makes the analogy to
the Ising model then one can write down the free energy {\it per plaquette}
\cite{ShiWortis}.  This is done most easily by using a transfer matrix
\cite{Huang}.  However, to make
connections to free energy barriers, we find it most natural to compute the
free energy {\it per unit projected length along one of the lattice axes}.

The calculation proceeds as in Sec.\ II of Ref.\ \cite{OurLongPaper}, except
with the addition of a field term $-m_i {\tilde h}$ to the energy
of a column in Eq.\ (2.10) of \cite{OurLongPaper}.  Here, we are interested
only in the large--size ($L \to \infty$)
limit in which we can write the partition function as
\begin{equation}
        Z(L) = e^{-4J_2 L/T} \Biggl [1+{e^{-(8J_2-\tilde h)/T}\over{
		1 - e^{-(4J_2-\tilde h)/T}}} \Biggr ]^L
        \ ,
        \label{partitionfunction}
\end{equation}
where $L$ is the projected length along the [010] axis.
The quantity $f_p(T,\tilde h)$, a free energy per unit projected length at a
given applied ``field" $\tilde h$ is then
\begin{eqnarray}
        f_p(T,\tilde h) &&= -{T\over{L}} \ln[Z(L)] \nonumber\\
			&&= 4 J_2 - T \ln \biggl ( 1 + {q^2 H\over{1-qH}}
				\biggr )
        \ ,
        \label{barrierperlength}
\end{eqnarray}
where $q \equiv \exp(-4 J_2 /T)$ and $H \equiv \exp(\tilde h/T)$.
The (average) tangent of the angle of the step with respect to the
[010] axis is given by
\begin{eqnarray}
	\tan(\theta) &&= -{\partial f_p(T,\tilde h)\over{\partial {\tilde h}}}
				\nonumber\\
		     &&= {q^2 H\over{(1 - qH)(1 - qH + q^2 H)}}
		     \ .
        \label{stepangle}
\end{eqnarray}
This equation can be
inverted in order to express $\tilde h$ as a function of
$r \equiv \tan(\theta)$:
\begin{equation}
        {\tilde h} = T\ln \Biggl
			[{(2-q)r+q-\sqrt{(r^2+1)q^2 +
			2 (2-q)q r}\over{2q(1-q)r}} \Biggr ]
        \ .
        \label{field}
\end{equation}

The step free energy as a function of angle is then given by
\begin{equation}
        f_s(T,\theta) =
		[f_p(T,\tilde h) + {\tilde h}\tan(\theta)] \cos(\theta)
        \ ,
        \label{stepfreeenergy}
\end{equation}
where, for a given angle $\theta$, ${\tilde h}$ is determined by (\ref{field}).
The addition of ${\tilde h}\tan(\theta)$ takes the
Legendre transform of $f_p(T,\tilde h)$ to give us $f_p(T,\theta)$.
The physical interpretation
of this transform is simply that this term subtracts off, from
$f_p(T,{\tilde h})$, the energy contribution due to the coupling of the field
to
the angle, thus leaving us with what the free energy of a step at angle
$\theta$ would be in the absence of any applied ``field."
The factor $\cos(\theta)$ is necessary to give us the free energy per
unit (macroscopic) step length, $f_s$, rather than the free energy per unit
projected length, $f_p$.

There are two ways in which the calculation above is
useful to us.  First, for $T < T_{CR}$ and $\Delta\mu = 0$, the free energy
barrier per unit length for shrinking of a cubical projection of edge length
$L$ (in the large $L$ limit)
is given rigorously by $f_B(T) \equiv f_p(T,\tilde h = 0)$.
This is also, roughly speaking, the appropriate barrier
to consider for a coarsening surface (as discussed in Sec.\ \ref{sec3a}),
although the value of $f_B(T)$ can no longer be so rigorously determined.
In particular, it would depend on precisely how we choose to define the
characteristic length scale $L$.  For coarsening in the case $\Delta\mu \ne 0$,
the important quantity that enters into the equations
is $f_p(T,\theta=\pi/4)$ as discussed in Sec.\ \ref{sec4}.

Second, we can extend the above results to derive the ``opening angle"
$\theta_c(T)$, discussed in Sec.\ \ref{sec3b}, in the temperature regime
$T_{\rm CR} < T < T_{\rm ER}$.
To determine this angle, we consider a surface $[hkl]$ with $0 \le k \le h$ and
$0 \le l \ll h$.
Such a surface is thus very close to being a $[hk0]$ surface and if
it is unstable then (independent of which of the conical
point or ridge scenarios discussed in Sec.\ \ref{sec3b} is correct),
it will break up (at least primarily) into surfaces which are vicinal to
the [100] orientation and surfaces which are vicinal to the [010]
orientation.  These surfaces will have steps across them of a definite angle,
given by $\theta_c(T)$.  This opening angle is then
determined by the condition that it minimize the total free energy of the
resulting surface subject to the constraint that
the surface has the correct average tilt in the $\hat z$ direction.
The free energy of the surface is given by $f_s(T,\theta) L_s$ where $L_s$ is
the total length of step.  The tilt is given by $L_s \sin(\theta)$.  Therefore,
$\theta_c(T)$ is that angle $\theta$ which minimizes the quantity
$f_s(T,\theta)/\sin(\theta)$ \cite{Wrong}.
This one--dimensional minimization can easily be performed numerically.
The results are shown in Fig.\ \ref{Angle}.



\begin{figure}
\caption[]{The decomposition of a surface in the [111]--RSOS model
for the Hamiltonian of Eq.\ (\ref{hamiltonian}) at $T < T_{\rm CR}$.
Shown is a [111] surface which has been quenched from infinite temperature to
$T = 3 J_2$ at times (a) $t = 0$, (b) $t = 100$, and (c) $t = 10000$
(in MC steps/plaquette) following the quench.
}
\label{surface}
\end{figure}

\begin{figure}
\caption[]{Qualitative thermal evolution of the equilibrium crystal shape
(ECS) for the model specified by the Hamiltonian of Eq.\ (\ref{hamiltonian}),
after Ref.\ \cite{RottmanWortis}.
}
\label{ECS}
\end{figure}

\begin{figure}
\caption[]{The stable surface orientations for the model specified by the
Hamiltonian of Eq.\ (\ref{hamiltonian}) in various temperature regimes.
This figure shows an eighth of a sphere, representing surface orientations
by the unit vector normal to the surface.
The shaded regions on the figure show those surface orientations which
appear on the ECS of Fig.\ \ref{ECS}.  All surface orientations
in the unshaded regions are thermodynamically unstable \cite{Herring} and
surfaces prepared in such orientations will decompose into some combination
of stable surfaces.
}
\label{AllowedOrientations}
\end{figure}

\begin{figure}
\caption[]{Equilibrium configurations for a [15,5,1] surface in
the [111]--RSOS model at various temperatures: (a) $T = 4 J_2$,
(b) $T = 14 J_2$, and (c) $T = 34 J_2$.  Minimization of the total surface
free energy, under the assumption that the surface is close enough to being
vicinal that the steps do not interact, gives
$T_{\rm ps}([15,5,1]) \approx 22.6 J_2$ for the phase separation temperature
(See appendix and Fig.\ \ref{Angle}).
This estimate should provide an upper bound for the actual value of
$T_{\rm ps}([15,5,1])$.
}
\label{Separated}
\end{figure}

\begin{figure}
\caption[]{The ``ridge scenario."  In (a), we show a closeup of the corner
region of the ECS for $T_{\rm CR} < T < T_{\rm ER}$,
under the assumption that the sharp edges between the facets continue
into the curved part of the ECS, finally terminating at second--order critical
points.  At any point along this edge, there is ``coexistence" between
two surface orientations.  (Here, sharp first--order edges are
shown by dashed lines while the second--order Pokrovsky--Talapov boundaries
between the facets and the curved part of the ECS are shown by solid lines.
The solid circles represent second--order critical points at the end of the
first--order lines.)  In (b), we present the diagram of the stable
surface orientations, showing how any unstable surface will phase separate
into the two coexisting stable surfaces.  In our example, any orientation along
the dashed curve (such as the one indicated by the open circle) decomposes into
the two (marginally) stable surface orientations indicated by the solid
circles.
}
\label{ridge}
\end{figure}

\begin{figure}
\caption[]{The ``conical point scenario."  In (a), we show the
corner region of the ECS for $T_{\rm CR} < T < T_{\rm ER}$, under
the assumption that the sharp edge between the facets
terminates at the point where it meets the curved region of the ECS.
At this ``conical point" (indicated by a solid circle), an entire
one--parameter family of surface orientations coexist.  In the curved part of
the crystal shape, these orientations come together at this point just as they
do at the tip of a cone.
In (b), we present the diagram of the stable surface orientations, showing
how an arbitrary unstable surface orientation (shown by an open circle) will
decompose into a combination of the entire one--parameter family of coexisting
surface orientations (shown by the bold solid curve).
}
\label{conicalpoint}
\end{figure}

\begin{figure}
\caption[]{Phase separation temperature $T_{\rm ps}(\theta)$ for a vicinal
surface
as a function of the angle of the steps across it in the [111]--RSOS model,
as calculated in the appendix.  Alternatively, as explained in the text,
the inverse function $\theta_c(T)$ can be thought to represent
the angle of the steps across the vicinal surfaces which an unstable crystal
surface $[hkl]$ with $0 \le k \le h$ and
$0 \le l \ll h$ decomposes into.  The symbols represent
measurements of $\theta_c(T)$ from simulations of this decomposition process.
For a finite value of $J_1/J_2$ ({\it i.e.}, relaxing the RSOS constraint),
the curve would look qualitatively similar for large angles but would not
diverge as $\theta \to 0$.  Instead, we would have
$T_{\rm ps}(\theta \to 0) = T_{\rm ER}$.
}
\label{Angle}
\end{figure}

\begin{figure}
\caption[]{An equilibrium configuration for a [13,13,4] surface in
the [111]--RSOS model at $T = 8 J_2$.  Note that the surface appears to have
phase--separated, but the equilibrium state is quite complicated.  This may
be an indication that many different surface orientations are
involved, as one would expect from the conical point scenario.
}
\label{NotVicinal}
\end{figure}

\begin{figure}
\caption[]{Shrinking a cubical projection in the [111]--RSOS model.
(a) shows a cubical projection with $L = 4$. (b) shows the same
projection once one face has been removed.  For $\Delta\mu \ge 0$, this
configuration is lower in energy than that of (a) by $4 J_2 + \Delta\mu L^2$.
}
\label{CubicalProjection}
\end{figure}

\begin{figure}
\caption[]{The time to shrink a cubical projection for $\Delta\mu = 0$.
Shown are the results of
Monte Carlo simulations for the average time to remove the
first face from a cubical projection of size $L$ [such as that shown in
Fig.\ \ref{CubicalProjection}(a)] for the case $\Delta \mu = 0$.
Each point is an average over 900 runs
with standard error smaller than the symbol size.
The dotted lines are one--parameter fits to the
form $t = t_0(L) e^{F_B(L,T=0)/T}$ with $F_B(L,T=0)$ given by
Eq.\ (\protect{\ref{BarrierNo}}) and $t_0(L)$ a free parameter.
Between 4 and 15 of the lowest temperature
data points are used for each fit.
}
\label{ShrinkingProjection}
\end{figure}

\begin{figure}
\caption[]{The time to shrink a cubical projection for nonzero $\Delta\mu$.
Shown are the results of Monte Carlo simulations for the
average time to remove the
first face from a cubical projection of size $L$ for the case
$\Delta \mu/J_2 = 2$.  Each point is an average over 900 runs
with standard error smaller than the symbol size.
The dotted lines are one--parameter fits to the
form $t = t_0 e^{F_B(L,T=0)/T}$ with $F_B(L,T=0)$ given by
Eq.\ (\ref{BarrierMu})
and $t_0(L)$ a free parameter.  Between 6 and 18 of the lowest temperature
data points are used for each fit.  For $L \le 4$, some of the fitted data
lies off scale.
}
\label{ShrinkingProjectionMu}
\end{figure}

\begin{figure}
\caption[]{Growth of the characteristic length scale $L(t)$ during
coarsening of a [111] surface at $T = 2 J_2$ for various values of
$\Delta\mu$ (given on the figure in units of $J_2$).  The runs were
performed on a $120^2$ system.  For each value of $\Delta\mu$,
the Monte Carlo data has been averaged over
77--150 runs, with error bars showing the standard error.
A line of slope 1/4 has been shown to facilitate comparison
with the power law $L(t) \sim t^{1/4}$.
The characteristic time when we would expect a change in growth law on the
basis of Eq.\ (\protect{\ref{CharacteristicTime}}) and using
$f_p(T=2 J_2,\theta=\pi/4) \approx 6.75 J_2$,
is $t_c \approx 3\times 10^2$, $1\times 10^4$,
$2\times 10^6$, and $8\times 10^9$ for
$\Delta\mu = 2.0$, $1.2$, $0.8$, and $0.5$, respectively.
}
\label{Coarsening}
\end{figure}


\begin{references}
\bibitem{Elastic} Another possibility which has been discussed
(Ref.\ \cite{Phaneuf1,Stewart,Fong,Alerhand,Marchenko})
is that elastic stresses set an upper bound on the size of the coarsening
facets.
\bibitem{RoughlySpeaking}  Technically speaking, one should worry about the
fact
that the electrostatic interaction between ions
would be expected to be long--ranged.
However, this problem can be avoided by considering the elementary units to
be cubes of 8 atoms which again sit on a cubic lattice (with twice the
atomic lattice spacing) and have only octopolar interactions between them.
For details, see \cite{ShiWortis}.
\bibitem{CrystalographicNotation}
We use the notation $\{100\}$ to refer to the set of surface orientations
which includes the $[100]$ and all symmetrically--equivalent orientations.
\bibitem{RoughSense}  It is tempting to claim that the fact that $T_{\rm CR}$
is
above room temperature explains the sharp cubical shapes seen in ordinary
grains of table salt.  However, such grains were formed by grinding and
are not in equilibrium with their vapor.  Thus, while one might make the
general
point that these grains show the basic preference of NaCl to expose its
$\{100\}$ facets, any attempt to make a more detailed correspondence
between equilibrium shapes and table salt should probably be ``treated
with a grain of salt."
\bibitem{ConfusingNotation} We use the same symbol $h$ both here
and in expressing surface orientations as $[hkl]$, but trust that our
meaning is always clear from context.
\bibitem{SurfaceTension} In Ref.\ \cite{SixVertex}, we noted how the
projected surface tension has the linear form $F(x,y) = A |y-x|$
in the coexistence regions if there is a conical point.  If, instead, we had
the ridge scenario, the projected surface tension would be of the form
$F(x,y) = f(|y-x|)$ in the coexistence regions, where $f(z)$ is a function
with positive first and second derivatives.
\bibitem{Unphysical}  This FCC model, however, is likely less physically
relevant since the interactions do not have a simple interpretation in terms of
real materials:  Because some of the nearest--neighbors of the
nearest--neighbors of
a lattice site A are themselves nearest--neighbors of site A,
there is no way to decorate the sites of an FCC lattice with atoms in such a
way that the nearest--neighbors of a given atom are all of unlike species while
the next--nearest--neighbors are all of like species.
\bibitem{SixVertexScenerio}  Curiously enough, the original explanation of the
solution to the six--vertex model by Sutherland, Yang, and Yang
\cite{SixVertexSolution} has an incorrect phase diagram which corresponds to
neither the ridge nor the conical point scenarios.  Instead, their phase
diagram
corresponds to having all surface orientations stable at any nonzero
temperature, except perhaps for those marginal orientations parallel to the
edge!
\bibitem{Wrong} Note that $\theta_c(T)$ is {\it not} determined
simply by the condition $f_s(T,\theta) = 0$, as one would find if one simply
compared the surface's free energy to that of a surface which decomposes into
the $[100]$, $[010]$, and $[001]$ surface orientations.  This is because
even a surface which cannot lower its free energy by decomposing into these
facet orientations may still be able to lower its free energy by breaking up
into other surface orientations.
\end{references}
\end{document}